\documentclass[aps,showpacs,showkeys]{revtex4}
\usepackage{amsmath, amsthm, amssymb,epsfig,alltt}

\newcommand{\nn}{\nonumber }

\def\a{\alpha}
\def\b{\beta}

\def\p{\partial}
\def\m{\mu}

\def\t{\tau}

\def\s{\sigma}

\def\nn{\nonumber}

\def\2pap{2\pi\alpha^\prime}

\def\beq{\begin{eqnarray}}
 \def\eeq{\end{eqnarray}}
 \def\4pap{4\pi\a^\prime}

 \def\ap{{\a^\prime}}

 \def\tal{{\tilde \alpha}}

\begin{document}


\title[Short Title]{Quantum Brownian Motion on a Triangular Lattice and Fermi-Bose Equivalence:\\
An Application of Boundary State Formulation}

\author{Taejin Lee}
\email{taejin@kangwon.ac.kr}

\affiliation{
Department of Physics, Kangwon National University,
Chuncheon 200-701, Korea}

\affiliation{
Pacific
Institute for Theoretical Physics Department of Physics and
Astronomy, University of British Columbia, 6224 Agricultural
Road, Vancouver, British Columbia V6T 1Z1, Canada}

\date{\today}

\begin{abstract}
We discuss the Bose-Fermi equivalence in the quantum Brownian motion (QBM) on a triangular lattice, mapping the action for the QBM into a string theory action
with a periodic boundary tachyon potential. 
We construct new Klein factors which are more appropriate 
than the conventional ones to deal with the 
quantum field theories defined on a two dimensioanl space-time with boundaries.
Using the Fermi-Bose equivalence with the new Klein factors, we show that the 
model for the quantum Bownian motion on a triangular lattice is equivalent to the 
Thirring model with boundary terms, which are quadratic in fermion field operators, in the off-critical regions and to a $SU(3)\times SU(3)$ free fermion theory with quadratic boundary 
terms at the critical point. 
\end{abstract}

\maketitle

\section{Introduction}

In two dimensional space-time some boson theories can be mapped into corresponding fermion theories, and fermion theories vice versa . The relation between the field operators of the boson theories and the fermion theories is known as the Fermi-Bose 
equivalence \cite{Coleman,stone}, which has been an essential tool to study various
important models in particle physics, string theory and condensed matter physics.
Often the fermionized or the bosonized theory turns out to be much easier to treat
than the model which we begin with. Sometimes we obtain exact results simply by fermionizing or bosonizing the models. A recent example is the rolling tachyon \cite{sen02r,senreview,rollingmore}, which describes the time evolution of unstable D-branes in string theory. The rolling tachyon is described by a string theory on a disk with a boundary potential due to the tachyon condensation on a D-brane. If a Wick rotation to the Euclidean time is taken, the action for the rolling tachyon becomes a string theory action on a disk with a marginal periodic potential. This action corresponds to 
the one dimensional quantum Brownian motion (QBM) action \cite{caldeira83phy,fisher} 
at its critical point. The marginal periodic potential can be rewritten as 
a boundary mass term for a fermion, which is quadratic in the fermion field, if 
the theory is fermionized. Thus, the theory becomes exactly solvable. An explicit 
expression of the boundary state for the rolling tachyon has been given in refs.\cite{Tlee:2005ge,Hassel}. 

If we consider a D-brane of the bosonic string theory with one direction wrapped
on a circle, we may find a tachyon mode coming from the first momentum mode of the standard tachyon along the circle direction, which leads us to an inhomogeneous 
rolling tachyon. In condensed matter physics, this inhomogeneous rolling tachyon 
corresponds to the spin-dependent Tomonaga-Luttinger model with a scattering potential,
which describes quantum transport through a single barrier \cite{furusaki}.
The critical behavior of the model has be studied in ref.\cite{Tlee:2005ge,Hassel}, 
applying the fermionization and the boundary state formulation. The inhomogeneous rolling tachyon can be mapped onto a Thirring model with two flavours with a boundary mass if fermionized. At the critical point the model reduces to a $SU(2) \times SU(2)$ fermion theory with a boundary mass term, which is exactly solvable.

The model we will discuss in this paper, 
utilizing the Fermi-Bose equivalence and the boundary state formulation, is the QBM \cite{kane92,yi,affleck00} on a triangular lattice. 
The action for the model is equivalent to a Polyakov string action on a disk with a periodic boundary potential on a triagular lattice. 
In order to fermionize the model we need to introduce three species fermion fields and appropriate Klein factors, which ensure the fundamental anti-commutation relations between the fermion fields. Here, it should be noted that the construction of Klein factors for the model
with three species fermion fields is much more involved than the cases of model with one or two 
species of fermion fields. In this paper we will elaborate the construction to provide
an explict expression for Klein factors.

There are two conventional methods to construct the Klein factors \cite{senechal}: 
The first one is to 
introduce additional anticommuting factors, which obey the Clifford algebra. They are 
usually represented by direct products of Pauli matrices. It requires an extra Hilbert space on which the Klein factors act.
The second one is to make use of the zero modes of the momentum fields, which are conjugate to the boson fields. One may find that exponentials of the zero modes of the momentum fields anti-commute the exponentials of the boson fields, 
using the Baker-Campbell-Hausdorff formula. Both methods work equally well for the theories defined on the two dimensional space-time without a boundary.
If the two dimensional space-time on which the theory is defined has a boundary,
we need to refine the conventional methods to construct the Klein factors. 
It is desirable that the boundary conditions for the simple boundary states such as the Neumann boundary state and the Dirichlet boundary state are represented linearly in
terms of both boson fields and fermion fields only. However, with the conventional Klein factors, it seems difficult to construct such simple boundary states for the systems.
If we adopt the conventional methods, the Klein factors enter into the boundary condtions
as we will see shortly. 

We will derive all the necessary conditions for the Klein 
factors not to enter into the fermionized action 
and show that there exist some solutions to them by an explicit construction.
The newly constructed Klein factors help us to analyze the critical behavior of the model, since
with them the QBM model can be mapped into a Thirring model with three flavors.
At the critical point the QBM model can be mapped into a $SU(3) \times SU(3)$ free fermion theory with quadratic boundary terms at the critical point. Hence, it becomes an exactly solvable model at the critical point.

\section{The Quantum Brownian Motion on a Triangular Lattice}

The Brownian motion of a classical particle on a triangular lattice is described by a simple Langevin equation which contains a frictional force and a conservative force due to a periodic potential on the lattice. The frictional force may be produced by coupling 
the particle linearly to a bath or an environment which consists of an infinite set of Harmonic oscillators as Caldeira and Leggett \cite{caldeira83ann} suggested. If we integrate out the infinite degrees of freedoms of the bath, we obtain the Euclidean action for the QBM as follows 
\beq \label{action1}
S_{QBM} &=& \frac{\eta}{4\pi} \int^{\beta}_{0} dt dt^\prime 
\frac{\left({\bf X}(t) - {\bf X}(t^\prime)\right)^2}{(t-t^\prime)^2} 
+ \frac{M}{2} \int^{\beta}_{0} dt \dot {\bf X}^2 + V_0\int^{\beta}_{0} dt \sum_{i=1}^3 \cos \left({\bf k}_i \cdot {\bf X}\right) 
\eeq
where $\beta=1/T$ and 
\beq
{\bf k}_1 = (\frac{1}{2},\frac{\sqrt{3}}{2}), ~~~
{\bf k}_2 = (\frac{1}{2},-\frac{\sqrt{3}}{2}), ~~~
{\bf k}_3 = \left(-1,0 \right).
\eeq
The two dimensional vector ${\bf X} = (X^1, X^2)$ describes the trajectory of the Brownian particle as a function of the imaginary time.
In Eq.(\ref{action1}) $\eta$ is the frictional constant which measures the strength of the coupling to the bath and $M$ is a mass for the Brownian particle. 
The third term denotes the periodic potential with a dimensionless coupling constant $V_0$.
Throughout this paper, we set $\hbar = k_B =1$. 

The first term is the non-local friction term, which results from the integration over the infinite degrees of freedoms of the bath in the path integral. It is due to this 
non-local interaction that the QBM exhibits non-trivial a phase transition despite being 
a quantum mechanical model. Since we are only interested in the long-time behavior of the system, we may ignore the kinetic term, which only plays a role of regulator in the long-time analysis. 

Identifying the time dimension as the boundary of a disk diagram, we may map
this model into a string theory action with the periodic boundary potential on a triangular lattice
\beq \label{stringaction}
S = \frac{\eta}{4\pi} \int d\t d\s \p_\a {\bf X} \cdot \p^\a {\bf X} + \frac{V}{2}
\int d\s \sum_{i=1}^3 \left(e^{i{\bf k}_i \cdot {\bf X}}+ e^{-i{\bf k}_i \cdot
{\bf X}} \right) 
\eeq
where $\s= \frac{2\pi}{\b} t$ and ${\bf X} (\t=0,\s) = {\bf X}(t)$ and $V= {V_0}\frac{\beta}{2\pi}$. We see that the slope parameter $\a^\prime$ of the corresponding string theory is $1/\eta$. 
The non-local action may be obtained as 
we integrate the bulk degrees of freedom of ${\bf X}(\t,\s)$ out in the path integral for the partition function, leaving the boundary degrees of freedom ${\bf X}(\t=0,\s)$ only.
The bulk degrees of freedom of ${\bf X}$ play the role of the bath.

We may embed the 
two dimensional model into the three dimensional one, introducing a free auxiliary field $X^3$
\beq
\frac{1}{4\pi \ap} \int d\t d\s \sum_{a=1}^2 \p_\a {X^a} \p^\a {X^a} \rightarrow
\frac{1}{4\pi \ap} \int d\t d\s \sum_{a=1}^3 \p_\a {X^a} \p^\a {X^a} 
\eeq
where $\a^\prime = 1/\eta$. 
Since the free field $X^3$ does not appear in the boundary interaction terms, its contributions to the partition function and the expectation values of physical operators can be easily factored out. As we shall see, with inclusion of the auxiliary field $X^3$, the underlying symmetry of the model, which turns out to be $SU(3) \times SU(3)$ at the critical point, becomes manifest, if the model is correctly fermionized.  

In order to fermionize the system it may be convenient to rewite the action in terms of
$(\phi^1, \phi^2, \phi^3)$ which is related to $(X^1, X^2, X^3)$ by an $O(3)$ rotation
\beq
\phi^1 &=& \frac{1}{\sqrt{2}} X^1 
+ \frac{1}{\sqrt{6}} X^2 + \frac{1}{\sqrt{3}} X^3, \nn\\
\phi^2 &=& -\frac{1}{\sqrt{2}} X^1 
+ \frac{1}{\sqrt{6}} X^2 + \frac{1}{\sqrt{3}} X^3, \\
\phi^3 &=& -\sqrt{\frac{2}{3}} \,\,X^2 + \frac{1}{\sqrt{3}} X^3. \nn
\eeq
If we rewrite the action for QBM in terms of $\phi^a,~~ a =1 ,2 ,3$, we have
\beq
S = - \frac{1}{4\pi \ap} \int d\t d\s \p_\a \phi^a \p^\a \phi^a + \frac{V}{2}
\int d\s \sum_{a=1}^3 \left(e^{\frac{i}{\sqrt{2}}\left(\phi^a-\phi^{a+1}\right)}
+e^{-\frac{i}{\sqrt{2}}\left(\phi^a-\phi^{a+1}\right)} \right)
\eeq
where $\phi^{a+3} = \phi^a$.

\section{Boundary State and Fermionization}

We consider the critical case where $\ap= \eta = 1$ first and will return to the general case later. The most efficient way to calculate the partition function and correlation 
functions of physical operators is to construct a boundary state. Since, moreover 
at the critical point the boundary potential may becomes quadratic in fermion fields if the model is appropriately fermionized, we may calculate the exact form of 
the boundary state. Thus, the partition function and other correlation functions of 
physical operators may be evaluated exactly at the critical point if the Fermi-Bose 
equivalence is used. To be explicit, if once the boundary state $\vert B_{QBM} \rangle$ is constructed, the partition function for the QBM may be written as 
\beq
Z_{QBM} &=& \langle 0 \vert B_{QBM} \rangle = \int D[{\bf X}] \exp\left(-S_{QBM}\right) ,
\eeq
and the correlation functions of physical operators in terms of the boundary state as 
\beq
\langle O(t_1) \cdots O(t_n) \rangle = \langle 0 \vert \langle O(\s_1) \cdots O(\s_n)
|B_{QBM} \rangle
\eeq
where $\s_i = 2\pi t_i/T$ and $|0\rangle$ is the ground state of the corresponding 
string theory.

If the periodic boundary interaction is absent, the boundary conditions for $\phi^a$,
$a = 1, 2, 3$,  
would be Neumann: $ \left(\phi^a_L - \phi^a_R\right)\vert_{\s = 0} = 0$. Thus, the boundary state for DHM may be written formally as 
\beq
|B_{QBM}\rangle = :\exp\left[-\frac{V}{2}
\int d\s \sum_{a=1}^3 \left(e^{\frac{i}{\sqrt{2}}\left(\phi^a-\phi^{a+1}\right)}
+e^{-\frac{i}{\sqrt{2}}\left(\phi^a-\phi^{a+1}\right)} \right)\right]:|N,N,N\rangle
\eeq
where
$\left(\phi^a_L- \phi^a_R\right)|N,N,N\rangle = 0$.
Hence we may write the boundary state and the boundary interaction as 
\beq 
|B_{QBM}\rangle &=& :\exp\left[-\frac{V}{2}
\int d\s \sum_{a=1}^3 \left(e^{i \sqrt{2}\left(\phi^a_L-\phi^{a+1}_L\right)}
+e^{-i\sqrt{2}\left(\phi^a_L-\phi^{a+1}_L\right)} \right)\right]:|N,N,N\rangle \\
&=& :\exp\left[-\frac{V}{2}
\int d\s \sum_{a=1}^3 \left(e^{i \sqrt{2}\left(\phi^a_R-\phi^{a+1}_R \right)}
+e^{-i\sqrt{2}\left(\phi^a_R-\phi^{a+1}_R\right)} \right)\right]:|N,N,N\rangle \nn
\eeq
where $V$ may be renormalized. (The boson field $\phi$ may be written as a sum of the 
left moving $\phi_L$ and the right moving $\phi_R$ boson fields, defined as follows:
$ \phi_L ~=~
 \frac{1}{\sqrt{2}} x_L+ \frac{1}{\sqrt{2}} p_L \sigma+\frac{i}{\sqrt{2}}\sum_{n\neq
 0}\frac{\a_n}{n}e^{-ni\sigma}, ~~
 \phi_R ~=~
 \frac{1}{\sqrt{2}} x_R- \frac{1}{\sqrt{2}} 
 p_R \sigma+\frac{i}{\sqrt{2}}\sum_{n\neq
 0}\frac{\tilde\a_n}{n}e^{ni\sigma)}$.)

Since $e^{-i\sqrt{2} \phi^a_{L/R}}$ has the same conformal dimension as the fermion
field operator $\psi^a_{L/R}$,
the boundary interaction term may be equivalent to a sum of quadratic terms in fermion field operators. But, we must treat the Klein factors with care before setting up the correct Bose-Fermi equivalence. We will apply the conventional methods to constuct the 
Klein factors first and will point out that the conventional ones are not adequate to deal with the theories defined on the space-time with a boundary and to construct the 
boundary states for the corresponing theories. Let us consider the simplest case of 
a single free boson theory and the simple boundary state such as the Neumann boundary state. We may represent the fermion operators in terms of the boson field operators as 
\beq
\psi_L = \eta_L :e^{-i\sqrt{2} \phi_L}:, \qquad 
\psi_R = \eta_R :e^{i\sqrt{2} \phi_R}:.
\eeq
In order to ensure the anti-commutation relations between the fermion operators we may require that they satisfy the Clifford algebra
$ \{\eta_i, \eta_j\} = 2 \delta_{ij},  i, j = L, R, $
and they commute with the boson operators $\phi_{L/R}$. This is one of the conventional 
methods to construct the Klein factors. They are usually represented by the Pauli matrices. To explicit, $ \eta_L = \s_1,  \eta_R = \s_2$. The Neumann boundary condition is simply given in terms of boson field operators as 
\beq
\phi_L|N\rangle = \phi_R|N\rangle.
\eeq
But its fermion representation cannot be written in terms of the fermion fields only
\beq
\psi_L|N\rangle = \eta_L \eta_R \psi^\dagger_R|N\rangle = i \s_3 \psi^\dagger_R|N\rangle,
~~\psi^\dagger_L|N\rangle = \eta_L \eta_R \psi_R|N\rangle = i \s_3 \psi_R|N\rangle.
\eeq
The Klein factors enter into the boundary condition. The free boson field theory defined on a space-time with a boundary is not mapped onto a free fermion field theory by the Pauli matrix representation of the Klein factors. 

The other conventional representation of the Klein factors is to make use of the zero modes of the momenta, conjugates to the boson field operators $\phi_L$ and $\phi_R$;
$ \eta_i = e^{i\pi \sum_{j<i} p_j}$. For the single boson theory, we may write
\beq
\psi_L = e^{-i\sqrt{2} \phi_L}, ~~
\psi_R = e^{i \pi p_L} e^{i\sqrt{2} \phi_R}.
\eeq
The Neumann boundary condition is now expressed in term of the fermion field operators
as
\beq
\psi_L |N \rangle = e^{i\pi p_L} \psi^\dagger_R |N\rangle, ~~
\psi^\dagger_L |N\rangle = e^{-i\pi p_L} \psi_R |N\rangle . 
\eeq
Note that the Klein factors enter into the boundary condition as before. Thus, the free
boson theory on a disk is not simply mapped onto a free fermion theory, contrary to our expectation.

If we may refine the second method, we may achieve our goal to map the bosonic 
boundary conditions for the simple boundary states into the fermion boundary conditions
which are linear in the field operators without the non-trivial Klein factors. 
The case of the single boson system has been worked out in ref.\cite{Tlee:2005ge}. The most general 
forms of the Klein factors may be written as 
\beq  \psi_{L} = e^{- \frac{\pi i}{2}(\a^L p_L + \b^L p_R)}
e^{-\sqrt{2}i X_L}, ~~\psi_{R} &=& e^{\frac{\pi i}{2} (\a^R p_L + \b^R p_R)}
e^{\sqrt{2}i X_R}. 
\eeq  
Then we may derive the conditions for the Klein factors to satisfy, requiring that 
the boundary condition for the simple boundary states are represented linearly in 
the fermion field operators with other conditions such as the fundamental anti-commutation
relations between the fermion field operators. It is not difficult to find a solution to those conditions: 
\beq   
\psi_{L} = e^{-\frac{\pi i}{2} p_R} e^{-\sqrt{2}i X_L}, \quad
\psi_{R} = e^{-\frac{\pi i}{2} p_L} e^{\sqrt{2}i X_R}, 
\eeq
with which the Neumann boundary condition is linearly represented as 
\beq   
\psi_L|N\rangle = i \psi^{\dagger}_R |N\rangle,\quad
\psi_L^\dagger|N\rangle = i \psi_R|N\rangle.
\eeq
The Klein factors for the case of two boson models have been constructed \cite{Tlee:2005ge} and applied
to the rolling tachyon in string theory \cite{Hassel} and the inhomogenous rolling tachyon \cite{tlee08}.
In the next section we will extend it to the case of the QMB which requires three boson fields for its free fermion description.

\section{Klein Factors and Simple Boundary States}

The Klein factors for the fermion fields which correspond to the 
the three bosons fields of the QBM may be parametrized as 
\beq
\psi^a_L = e^{-\frac{\pi}{2} i \sum_b\left(
\a^L_{ab} p^b_L + \b^L_{ab} p^b_R \right)} e^{-\sqrt{2} i X^a_L},~~
\psi^a_R = e^{\frac{\pi}{2} i \sum_b\left(
\a^R_{ab} p^b_L + \b^R_{ab} p^b_R \right)} e^{\sqrt{2} i X^a_R}
\eeq
where $a, b = 1, 2, 3$. 
The left moving boson field operators and right moving ones may be expanded in terms of the oscillator modes as follows:
 \begin{subequations}
 \label{expan}
 \begin{eqnarray}
 X_L(\tau+i\sigma)~=~
 \frac{1}{\sqrt{2}} x_L-\frac{i}{\sqrt{2}} p_L(\tau+i\sigma)+\frac{i}{\sqrt{2}}\sum_{n\neq
 0}\frac{\a_n}{n}e^{-n(\tau+i\sigma)}, \label{expan:a} \\
 X_R(\tau-i\sigma)~=~
 \frac{1}{\sqrt{2}} x_R- \frac{i}{\sqrt{2}} 
 p_R(\tau-i\sigma)+\frac{i}{\sqrt{2}}\sum_{n\neq
 0}\frac{\tilde\a_n}{n}e^{-n(\tau-i\sigma)}. \label{expan:b}
 \end{eqnarray}
 \end{subequations}
with the non-vanishing commutators
\beq
 \left[ x_L,p_L\right]~=~i,~~~ \left[ x_R, p_R\right] ~=~ i,~~~
 \left[ \a_m, \a_n \right] ~=~ m\delta_{m+n},~~~ \left[
 \tilde\a_m, \tilde\a_n \right] ~=~ m\delta_{m+n}.
\eeq 
The anti-commutation relations between the fermion field operators
$\psi_{aL},~\psi_{aL},~\psi^{a \dagger}_L,~ \psi^{a \dagger}_R$, are ensured
if
\beq \label{anti}
e^{\frac{\pi i}{2}(\a^L_{ab}-\a^L_{ba})} = -1,~~
e^{\frac{\pi i}{2}(\b^R_{ab}-\b^R_{ba})} = -1,~~
e^{\frac{\pi i}{2}(\a^R_{ab}-\b^L_{ba})} = -1,~~{\rm for}~ a\not=b.
\eeq
We note that the anti-commutation relations between the fermion operators alone 
cannot fix the Klein factors. Additonal conditions would be obtained by requiring that
the simple boundary states should be linearly represented in terms of the fermion
fields and the interaction terms should be represented in terms of the fermion fields
only in the fermion theory.

\subsection{The Neumann Boundary State and Klein Factors}

The boundary conditions for simple boundary states such as $|N,N,N\rangle$ and $|D,D,D\rangle$ should be realized in terms of the fermion operators without the Klein factors. This requirement will impose some conditions for the Klein factors.
We begin with the boundary state $|N,N,N\rangle$.  
The boundary condition for the state $|N,N,N\rangle$  is given linearly in terms of the bosonic operator as
\beq   
X^a_L|N,N,N\rangle = X^a_R|N,N,N\rangle   \eeq
where $a = 1, 2, 3$, which can be read in terms of normal modes as
\beq  x^a_L|N,N,N\rangle= x^a_R|N,N,N\rangle, \quad 
p^a_L|N,N,N\rangle= - p^a_R|N,N,N\rangle,\quad 
\a^a_n|N,N,N\rangle= - \tal^a_{-n}|N,N,N\rangle. \nn  
\eeq
This condition can be realized linearly in terms of the fermion operators if the
following conditions are satisfied 
\beq \label{nnn}
\a^L_{ab} -\a^R_{ab}  - \b^L_{ab} + \b^R_{ab} = 0,
\eeq
since 
\beq
\psi^a_L|N,N,N\rangle 
&=& e^{\frac{\pi}{2}i \b^L_{aa}} \psi^\dagger_{aR}
e^{\frac{\pi}{2}i \sum_b \left(\a^R_{ab} - \b^R_{ab} -\a^L_{ab} + \b^L_{ab} \right)p^b_L} |N,N,N\rangle .
\eeq
Under the condition Eq.(\ref{nnn}) we write the Neumann boundary condition in the fermion
theory as
\beq
\psi^a_L|N,N,N\rangle = e^{\frac{\pi}{2}i \b^L_{aa}}\psi^\dagger_{aR}
|N,N,N\rangle, \quad
\psi^{a \dagger}_L|N,N,N\rangle
= e^{-\frac{\pi}{2}i(\a^L_{aa}-\b^L_{aa}+\b^R_{aa})}\psi^a_R
|N,N,N\rangle. \eeq
These fermion boundary conditions should be also consistent with the fundamental anti-commutation relations between the fermion 
field operators. It yields additional conditions
\beq \label{cons1} e^{\frac{\pi}{2}i(-\a^L_{aa}+2 \b^L_{aa} -\b^R_{aa})}= -1, \,\,\, {\rm for} \,\, \, a = 1, 2, 3.
\eeq


\subsection{The Dirichlet Boundary State and Klein Factors}

The boundary condition for $|D,D,D\rangle$
is given in terms of the bosonic operator as 
\beq
X^a_L |D,D,D\rangle = - X^a_R |D,D,D\rangle, ~~~ a = 1, 2, 3.
\eeq
If it is written in terms of normal modes,
\beq  x^a_L|D,D,D\rangle= -x^a_R|D,D,D\rangle, \quad 
p^a_L|D,D,D\rangle=  p^a_R|D,D,D\rangle,\quad 
\a^a_n|D,D,D\rangle=  \tal^a_{-n}|D,D,D\rangle. \nn  
\eeq
Since it may be written in terms of fermion operator as
\beq
\psi^a_L|D,D,D\rangle
&=& e^{-\frac{\pi}{2}i\left(\b^L_{aa}+ \b^R_{aa}\right)} \psi^a_R e^{-\frac{\pi}{2}i \sum_b\left(\a^L_{ab}+
\a^R_{ab} + \b^L_{ab} + \b^R_{ab}\right)p^b_L}|D,D,D\rangle. 
\eeq
We should impose the condition
\beq \label{ddd}
\a^L_{ab}+ \a^R_{ab} + \b^L_{ab} + \b^R_{ab} = 0.
\eeq
in order to have the Dirichlet boundary condition to be linearly represented
by the fermion field operators
\beq \label{ddd1}
\psi^a_L|D,D,D\rangle = e^{-\frac{\pi}{2}i\left(\b^L_{aa}+ \b^R_{aa}\right)} \psi^a_R|D,D,D\rangle .
\eeq
Under the condition Eq.(\ref{ddd}), we have
\beq \label{ddd2}
\psi^{a\dagger}_L|D,D,D\rangle 
= e^{-\frac{\pi}{2}i \left(\a^L_{aa}+ \b^L_{aa}\right)} \psi^{a\dagger}_R |D,D,D\rangle.
\eeq
These two boundary conditions Eq.(\ref{ddd1}) and Eq.(\ref{ddd2})
should be compatible with the fundamental fermion anti-commutation relations. It follows from this requirement that
\beq \label{cons2}
e^{-\frac{\pi}{2}i \left(\a^L_{aa} + 2\b^L_{aa} + \b^R_{aa}\right)} = -1 .
\eeq

{\bf Other boundary states and Klein factors}:
We may repeat the same procedure for other boundary states such as  $|D,N,N\rangle$, $|N,D,N\rangle$, $|N,N,D\rangle$,
$|D,D,N\rangle$, $|D,N,D\rangle$, $|N,D,D\rangle$. However, it does not produce any additional condition for the Klein factors.

{\bf Boundary interaction and the Klein factors}:
We may also apply the same procedure to the interaction terms $e^{i{\bf k}_i \cdot {\bf \Phi}} + e^{-i{\bf k}_i \cdot {\bf \Phi}}$, $i=1,2,3$. As we require that they are represented by
Hermitian quadratic fermion field operaters without nontrivial Klein factors. Some details of the
procedure are given in the appendix. The results are summarized as follows
\beq \label{interaction}
\a^{L/R}_{1a} - \a^{L/R}_{2a}-\b^{L/R}_{1a} + \b^{L/R}_{2a} &=& 0, \nn\\
\a^{L/R}_{2a} - \a^{L/R}_{3a}-\b^{L/R}_{2a} + \b^{L/R}_{3a} &=& 0, \\
\a^{L/R}_{3a} - \a^{L/R}_{1a}-\b^{L/R}_{3a} + \b^{L/R}_{1a} &=& 0, \nn
\eeq
where
$a = 1, 2, 3$.

\section{Solutions}

Now we will show that there exists solutions which satisfy all the conditions derived in the previous sections.  Making use of Eqs.(\ref{nnn}, \ref{ddd}), we may write
all the conditions only in terms of $\a's$
\beq \label{ab}
\b^L_{ab} = -\a^R_{ab}, ~~~ \b^R_{ab} = -\a^L_{ab}.
\eeq
Then the rest of equations Eq.(\ref{interaction}) read as 
\begin{subequations}
 \label{rest}
\beq
\a^L_{1b} + \a^R_{1b} -\a^L_{3b} -\a^R_{3b} &=& 0, \label{rest:a}\\
\a^L_{2b} + \a^R_{2b} -\a^L_{3b} -\a^R_{3b} &=& 0, \label{rest:b}\\
\a^L_{1b} + \a^R_{1b} -\a^L_{2b} -\a^R_{2b} &=& 0. \label{rest:c}
\eeq
\end{subequations}
Note that only two of these equations are independent.
We may replace these conditions by 
\beq \label{replace}
\a^L_{ab} + \a^R_{ab} = 0.
\eeq
We also note that if we make use of Eq.(\ref{ab}), the consistency conditions Eqs.(\ref{cons1},\ref{cons2}) reduce to
$e^{-\pi i \a^R_{aa}} = -1$,
which can be satisfied by 
\beq \label{cond1}
\a^{R}_{aa} = 2n^R_{aa} + 1, ~~~n^R_{aa} \in {\bf Z}, ~~~ a = 1, 2, 3
\eeq
The conditions Eq.(\ref{anti}) which ensure the anticommutation relations between fermion operators, $\psi^a_L$, $\psi^a_R$ reduce to 
\beq 
e^{\frac{\pi}{2}i(\a^L_{ab} -\a^L_{ba})} = -1, ~~
e^{\frac{\pi}{2}i(\a^R_{ab} + \a^R_{ba})} = -1  
\eeq
These conditions are satisfied by choosing 
\beq \label{cond2}
\a^L_{ab} -\a^L_{ba} = 2(2m^L_{ab}+1), ~~~
\a^R_{ab} + \a^R_{ba} = 2(2n^R_{ab}+1), ~~a < b, ~~m^L_{ab}, \, n^R_{ab} \in {\bf Z}.
\eeq
The remaining conditions for $\a^L_{ab}$ are
\begin{subequations}
\label{remain}
\beq
e^{\frac{\pi}{2}i(\a^L_{11} - \a^L_{13} - \a^L_{31} + \a^L_{33})} &=& 1, \label{remain:a}\\
e^{\frac{\pi}{2}i(\a^L_{22} -\a^L_{23} -\a^L_{32} + \a^L_{33})} &=& 1, 
\label{remain:b}\\
e^{\frac{\pi}{2}i(\a^L_{11}-\a^L_{12}-\a^L_{21}+\a^L_{22})} &=& 1.
\label{remain:c}
\eeq
\end{subequations}
With Eqs.(\ref{replace},\ref{cond1}) these conditions are satisfied by
\beq \label{cond4}
\a^L_{ab} +\a^L_{ba} = 2(2n^L_{ab}+1), \quad a<b, ~~ n^L_{ab} \in {\bf Z}.
\eeq
Since the Klein factors are not completely fixed yet by the conditions Eqs.(\ref{replace}, \ref{cond1}, \ref{cond2}, \ref{cond4}), we introduce further additional conditions
\beq \label{condadd}
\a^L_{12} = \a^L_{23} = \a^L_{31}.
\eeq
This condition guarantees that the boundary interaction terms have the same phase 
if they are fermionized.
A set of integers $n^L_{ab}, \, m^L_{ab}, \, m^R_{ab}$ with the additional conditions Eqs.(\ref{condadd}) determine an explicit representation of the Klein factors. The simplest one may be
\beq
\a^L = -\a^R = \b^L = -\b^R = 
\left(\begin{array}{rrr}
1 & 2 & 0 \\
0 & 1 & 2 \\
2 & 0 & 1 
\end{array}
\right).
\eeq
With this solution, we may write the boundary state for QBM at the critical point in fermion theory as 
\beq \label{boundstate}
|B_{QBM}\rangle &=& :\exp\left[ 
\frac{V}{2} \int d\s \sum_{a=1}^3 \left(\psi^{a\dagger}_L \psi^{a+1}_L - \psi^{a+1 \dagger}_L \psi^a_L\right) \right]:|N,N,N\rangle, \\
&=& :\exp\left[\frac{V}{2} \int d\s \sum_{a=1}^3 \left(\psi^{a+1 \dagger}_R \psi^a_R-
\psi^{a\dagger}_R \psi^{a+1}_R \right) \right]: |N,N,N\rangle, \nn
\eeq
where $\psi^{a+3}_{L/R} = \psi^a_{L/R}$.

\section{Conclusions}

The Fermi-Bose equivalence is one of essential tools 
to study the quantum field theories on two
dimensional space-time, which have various important applications in both string theory 
and condensed matter theory. But the Fermi-Bose equivalence has been discussed mostly for the theories on the space-time without a boundary. If the space-time has a boundary, we need to refine the conventional methods to construct the Klein factors, which guarantee the anti-commutation relations between the fermion field operators. We may take a single 
boson model as an example to show that the simple boundary conditions such as the Neumann and Dirichlet boundary conditions cannot be represented linearly in terms of the fermion field operators if the conventional methods are adopted. An explicit construction of more appropriate Klein factors for the single boson model has been
given in section III. 

In order to apply the Fermi-Bose equivalence to the QBM we need to 
generalize the new method to deal with the three boson model.
The most general form of the Klein factors for the three boson model has been constructed in section IV and all the necessary conditions for the Klein factors to satisfy have been derived. Explicitly solving the conditions, we show that there exists at least a set of solutions 
to those conditions. 
With the newly constructed Klein factors, the QBM model can be properly fermionized.  
At the critical point the QBM model can be mapped into a free fermion theory with quadratic boundary terms. It leads us to the exact boundary state for the QBM at the critical point Eq.(\ref{boundstate}). The constructed boundary state may be used to evaluate explicitly and 
exactly various physical quantities such as the partition function and the mobility. 

The QBM in the off-critical regions can be also fermionized by the improved Fermi-Bose 
equivalence. If $\eta=1/\a^\prime \not=1$, we may rewrite the bulk kinetic term as
\beq
\frac{1}{4\pi\ap} \int d\t d\s \p_\a \phi^a \p^\a \phi^a = \frac{1}{4\pi} \int d\t d\s \p_\a \phi^a \p^\a \phi^a  + \frac{1-\ap}{4\pi\ap} \int d\t d\s \p_\a \phi^a \p^\a \phi^a
\eeq
and treat the second term as an interaction. Applying the Fermi-Bose equivalence to the
QBM, we may have 
\beq
S_{QBM} = \frac{1}{2\pi} \int d\t d\s \sum_{a=1}^3 \left(\bar\psi^a \gamma^\mu \p_\mu \psi^a+ \frac{g}{4\pi} j^{\mu a} j_\m^a \right) + \frac{V}{4} \int d\s \sum_{a=1}^3
\left(\bar\psi^a \gamma^1 \psi^{a+1} - \bar \psi^{a+1} \gamma^1 \psi^a \right)
\eeq
where $g = \pi(1-\ap)/\ap$ and $j^{\mu a} = \bar\psi^a \gamma^\mu \psi^a$.
Thus, the QBM model is equivalent to a Thirring model \cite{Thirring,Klaiber} with
boundary terms, which are quadratic in fermion fields in the off-critical region. 
Using the Thirring action, we may easily calculate the radiative corrections 
to the boundary terms, which lead to the renormalization of $V$ \cite{tlee08}:
\beq
V = V_0 \left[1 + \frac{1-\ap}{2} \ln \frac{\Lambda^2}{\mu^2} \right]
= V_0 \left[\frac{\Lambda^2}{\mu^2}\right]^{\frac{(1-\ap)}{2}}.
\eeq
If $\ap <1$, the boundary terms become relevant operators and $V$ tends to grow while they become irrelevant operators and $V$ scales to zero when 
$\ap >1$ at low energy. 

At the critical point, the Thirring interaction vanishes and the model reduces to a $SU(3)\times SU(3)$ free fermion theory with some quadratic boundary terms. 
Since the boundary interactions are only quadratic in the fermion fields, we can 
solve the model exactly at the critical point. The partition function and other
physical correlation functions can be exactly calculated at the critical point if
the exact boundary state for the QBM Eq.(\ref{boundstate}) is employed.

The fermionized action for the QBM on a triangular lattice may be also useful to study the critical behaviors of the condensed matter theories \cite{affleck00,Affleck:1990by,yi,Oshikawa:2005fh}, which are closely related to the QBM. 
Since a honeycomb lattice consists of two triangular sublattices, it may not be difficult to
extend this work to the QBM on a honeycomb lattice, of which boson theory has been discussed 
by Yi and Kane in ref.\cite{yi}. The Fermi-Bose equivalance may help us to explore further the
critical behaviors of the QBM on a honeycomb lattice and may produce some exact results.
One more interesting avenue to explore, being equipped with the Fermi-Bose equivalne on
a triangular lattice (or on a honeycomb lattice), is the fully packed loop (FPL) model 
\cite{fpl1,fpl2,fpl3} on a honeycomb lattice. The effetive field theory for the fully packed loop model is similar to the local two dimensional action for the QBM \cite{referee}, Eq.(\ref{stringaction}). 
The only difference between two actions is that the periodic potential for the QBM is 
defined on the boundary while the periodic potential for the FPL model is defined
on the bulk. The Fermi-Bose equivalence, if properly extended to the
honeycomb lattice, would work equally well for the FPL model and help us to develop a 
new fermion field theory representation of the FPL. We may save it for a future work.
 
The QBM in one dimension corresponds to the rolling tachyon at the critical point \cite{Tlee:2005ge,Hassel} in string theory and the 
QBM on a two dimensional square lattice is equivalent to the inhomogeneous rolling tachyon
at the critical point \cite{tlee08}. 
It suggests that the QBM on a triangular lattice may be 
relevant to rolling tachyons in some compact target space-times. It is certainly an interesting task to explore this direction by extending this work.

\section*{Acknowledgement} 
Part of this work was done during the author's visit to UBC (Canada). He thanks G. Semenoff, P. Stamp, S. W. Lee and S. M. Ji for useful discussions. This work is supported 
in part by Research Institute of Basic Science, Kangwon National University.

\section*{Appendix}

\section*{Boundary Interaction and Klein Factors}

As we require that the boundary interaction terms are written as a sum of fermion bilinear field operators without nontrivial Klein factors, we obtain a set of conditions to be imposed. Since the boundary state for the system $|B_{QBM}\rangle$ can be obtained 
by applying the boundary interaction terms on the simple boundary state $|N,N,N\rangle$, 
we need to consider the fermionization of $\left(e^{i{\bf k}_i \cdot {\bf \Phi}} + e^{-i{\bf k}_i \cdot {\bf \Phi}}\right)|N,N,N\rangle, ~ i =1,2,3$.

\subsection{Boundary Interaction ($e^{i{\bf k}_1 \cdot {\bf \Phi}} + e^{-i{\bf k}_1 \cdot {\bf \Phi}}$) and Klein Factors}

The boundary interaction term
$e^{i{\bf k}_1 \cdot {\bf \Phi}}$ acting on the Neumann boundary state
may be written in terms of the left moving fermion field operators as
\beq
e^{i{\bf k}_1 \cdot {\bf \Phi}} |N,N,N\rangle &=&  e^{-\frac{\pi}{2}i\left(\a^L_{33} - \a^L_{13}\right)} \psi^\dagger_{1L} \psi_{3L} e^{-\frac{\pi}{2}i \sum_b \left(\a^L_{1b} - \a^L_{3b} -\b^L_{1b} + \b^L_{3b}\right)p^b_L}|N,N,N\rangle.
\eeq
It follows that the fermion form of the interaction term does not contain a non-trivial Klein factors if the following condition is satisfied
\beq \label{nnn1l}
\a^L_{1b} - \a^L_{3b} -\b^L_{1b} + \b^L_{3b} = 0, ~~ b = 1, 2, 3.
\eeq
Once this condition is imposed, the
boundary term $e^{i{\bf k}_1 \cdot {\bf \Phi}}$ may be written as 
\beq
e^{i{\bf k}_1 \cdot {\bf \Phi}} |N,N,N\rangle =e^{-\frac{\pi}{2}i\left(\a^L_{33} - \a^L_{13}\right)} \psi^\dagger_{1L} \psi_{3L}|N,N,N\rangle.
\eeq
The boundary term $e^{i{\bf k}_1 \cdot {\bf \Phi}}$ can be equally written in terms of the right chiral fermion field operators as
\beq
e^{i{\bf k}_1 \cdot {\bf \Phi}} |N,N,N\rangle 
&=&  e^{-\frac{\pi}{2} i(\b^R_{11} - \b^R_{31})} \psi^{\dagger}_{3R}
\psi_{1R} e^{\frac{\pi}{2}i \sum_b \left(\a^R_{3b} - \b^R_{3b} -\a^R_{1b} + \b^R_{1b} \right)p^b_L} |N,N,N\rangle. \nn
\eeq
Thus, if the following condition is satisfied
\beq \label{nnn1r}
\a^R_{3b} - \a^R_{1b} - \b^R_{3b} + \b^R_{1b} = 0,
\eeq
the the boundary term can be written as a bilinear fermion field operators
\beq
e^{i{\bf k}_1 \cdot {\bf \Phi}} |N,N,N\rangle = e^{-\frac{\pi}{2} i(\b^R_{11} - \b^R_{31})} \psi^{\dagger}_{3R} \psi_{1R} |N,N,N\rangle .
\eeq

As we repeat the same procedure for the boundary interaction term $e^{-i{\bf k}_1 \cdot {\bf \Phi}}$, we obtain the fermion form of the boundary interaction term as
\beq
e^{-i{\bf k}_1 \cdot {\bf \Phi}} |N,N,N\rangle = e^{-\frac{\pi}{2}i 
(\a^L_{11} - \a^L_{31})} \psi^\dagger_{3L} \psi_{1L} |N,N,N\rangle.
\eeq
A non-trivial Klein factor does not arise if the condition Eq.(\ref{nnn1l}) is satisfied.
Rewriting the boundary interaction term $e^{-i{\bf k}_1 \cdot {\bf \Phi}}$ in terms of the right moving fermion operators, we have
\beq
e^{-i{\bf k}_1 \cdot {\bf \Phi}} |N,N,N\rangle  = e^{\frac{\pi}{2}i \left(\b^R_{13}-\b^R_{33}\right)}\psi^\dagger_{1R} 
\psi_{3R}|N,N,N\rangle 
\eeq
under the condition Eq.(\ref{nnn1r}).

Therefore, we may write the boundary interaction term $e^{i{\bf k}_1 \cdot {\bf \Phi}} + e^{-i{\bf k}_1 \cdot {\bf \Phi}}$ in terms of the left moving fermion field operators as 
\beq
\left(e^{i{\bf k}_1 \cdot {\bf \Phi}} + e^{-i{\bf k}_1 \cdot {\bf \Phi}}\right) |N,N,N\rangle &=& 
\left(e^{-\frac{\pi}{2}i(\a^L_{33} - \a^L_{13})}\psi^\dagger_{1L} \psi_{3L}
+ e^{-\frac{\pi}{2}i(\a^L_{11} - \a^L_{31})} \psi^\dagger_{3L} \psi_{1L}\right) |N,N,N\rangle,
\eeq
or in terms of the right moving fermion field operators as
\beq
\left(e^{i{\bf k}_1 \cdot {\bf \Phi}} + e^{-i{\bf k}_1 \cdot {\bf \Phi}}\right) |N,N,N\rangle =
\left(e^{-\frac{\pi}{2}i(\b^R_{11} - \b^R_{31})}\psi^\dagger_{3R} \psi_{1R}
+ e^{\frac{\pi}{2}i(\b^R_{13} - \b^R_{33})} \psi^\dagger_{1R} \psi_{3R}\right) |N,N,N\rangle . 
\eeq
Note that, however, it is not manifestly Hermitian unless appropriate conditions for
the Klein factors are imposed. It can be achieved by introducing the conditions
\beq
e^{\frac{\pi}{2}i(\b^R_{11} - \b^R_{13} - \b^R_{31} + \b^R_{33})} = 1, ~~
e^{\frac{\pi}{2}i(\a^L_{11} - \a^L_{13} - \a^L_{31} + \a^L_{33})} = 1. 
\eeq

\subsection{Boundary Interaction $(e^{i{\bf k}_2 \cdot {\bf \Phi}} 
+ e^{-i{\bf k}_2 \cdot {\bf \Phi}})$ and Klein Factors}

We may rewrite the boundary interaction terms $(e^{i{\bf k}_2 \cdot {\bf \Phi}} + e^{-i{\bf k}_2 \cdot {\bf \Phi}})$ acting on $|N,N,N\rangle$ in terms of the left moving fermion field operators only as (without Klein factors)
\beq
\left(e^{i{\bf k}_2 \cdot {\bf \Phi}}+ e^{-i{\bf k}_2 \cdot {\bf \Phi}}\right) |N,N,N\rangle = 
\left(e^{-\frac{\pi}{2}i(\a^L_{22}-\a^L_{32})}\psi^\dagger_{3L} \psi_{2L} + e^{\frac{\pi}{2}i (\a^L_{23}
-\a^L_{33})} \psi^\dagger_{2L} \psi_{3L}\right) |N,N,N\rangle,
\eeq
if the following condition is satisfied
\beq \label{nnn2l}
\a^L_{3b} - \b^L_{3b} - \a^L_{2b} + \b^L_{2b} = 0.
\eeq
It can be made manifestly Hermitian if we require the following additional
condition
\beq
e^{\frac{\pi}{2}i(\a^L_{22} -\a^L_{23} -\a^L_{32} + \a^L_{33})}= 1.
\eeq

This boundary intraction term can be expressed equivalently in terms of the right moving fermion field operators as
\beq
\left(e^{i{\bf k}_2 \cdot {\bf \Phi}} + e^{-i{\bf k}_2 \cdot {\bf \Phi}}\right) |N,N,N\rangle =\left(e^{\frac{\pi}{2}i(\b^R_{23}
-\b^R_{33})} \psi^\dagger_{2R} \psi_{3R} + e^{-\frac{\pi}{2}i(\b^R_{22}
-\b^R_{32})} \psi^\dagger_{3R} \psi_{2R}\right) |N,N,N\rangle, 
\eeq
if the following condintion is satisfied
\beq \label{nnn2r}
a^R_{2b} - \b^R_{2b} -\a^R_{3b} + \b^R_{3b} = 0, ~~~ b = 1, 2, 3.
\eeq
If we further require that the interaction terms is manifestly Hermitian, we have
\beq
e^{\frac{\pi}{2}i(\b^R_{22} - \b^R_{23} -\b^R_{32} + \b^R_{33})} = 1.
\eeq

\subsection{Boundary Interaction $(e^{i{\bf k}_3 \cdot {\bf \Phi}} 
+ e^{-i{\bf k}_3 \cdot {\bf \Phi}})$ and Klein Factors}

Applying the same procedure repeatedly to the boundary intraction terms
$(e^{i{\bf k}_3 \cdot {\bf \Phi}} + e^{-i{\bf k}_3 \cdot {\bf \Phi}})$, we obtain the conditions for the Klein factors 
\beq \label{nnn3}
\a^L_{1b} -\a^L_{2b} -\b^L_{1b} + \b^L_{2b} = 0, ~~
\a^R_{1b} - \a^R_{2b} -\b^R_{1b} + \b^R_{2b} = 0, 
\eeq
under which the boundary interaction terms can be written in terms of
fermion field operators without non-trivial Klein factors as
\beq
\left(e^{i{\bf k}_3 \cdot {\bf \Phi}}+  e^{-i{\bf k}_3 \cdot {\bf \Phi}} \right)|N,N,N\rangle = \left(e^{-\frac{\pi}{2}i(\a^L_{11}-\a^L_{21})} \psi^\dagger_{2L} \psi_{1L} + e^{\frac{\pi}{2}i(\a^L_{12} -\a^L_{22})}\psi^\dagger_{1L} \psi_{2L} \right)
|N,N,N\rangle, 
\eeq
or
\beq
\left(e^{i{\bf k}_3 \cdot {\bf \Phi}}+ e^{-i{\bf k}_3 \cdot {\bf \Phi}}\right) |N,N,N\rangle = 
\left(e^{\frac{\pi}{2}i(\b^R_{12} -\b^R_{22})} \psi^\dagger_{1R} \psi_{2R}+                    e^{-\frac{\pi}{2}i(\b^R_{11}-\b^R_{21})}
\psi^\dagger_{2R} \psi_{1R} \right)|N,N,N\rangle .
\eeq
These boundary fermion interaction terms are manifestly Hermitian if 
\beq
e^{\frac{\pi}{2}i(\a^L_{11}-\a^L_{12}-\a^L_{21}+\a^L_{22})} = 1, ~~
e^{\frac{\pi}{2}i(\b^R_{11} -\b^R_{12} -\b^R_{21} +\b^R_{22})} = 1 .
\eeq


\begin{thebibliography}{0}


\bibitem{Coleman}
 S.~Coleman,
 Phys.~Rev.~ D {\bf 11}, 2088 (1975).
 
\bibitem{stone}
References in {\it Bosonization}, edited by M. Stone (World Scientific, Singapore, 1994)


\bibitem{sen02r} A.~Sen,
``Rolling tachyon", JHEP {\bf 0204}, 048  (2002) [hep-th/0203211];
``Time evolution in open string
theory,''  JHEP {\bf 0210}, 003 (2002) [hep-th/0207105];``Time and tachyon" [hep-th/0209122]; ``Open and closed strings from unstable D-branes, Phys. Rev. {\bf D68} (2003) 106003 [hep-th/0305011]; ``Dirac-Born-Infeld Action on the
Tachyon Kink and Vortex", [hep-th/0303057]; ``Open-Closed duality at tree level",
Phys. Rev. Lett. {\bf 91} (2003) 181601 [hep-th/0306137]; ``Open-Closed duality: Lessons from 
matrix model", Mod. Phys. Lett. {\bf A19} (2004) 841 [hep-th/0308068];``Rolling tachyon boundary
state, conserved charges and two dimensional string theory", JHEP
{\bf 0405}, 076 (2004), [hep-th/0402157]

\bibitem{senreview} See for a recenet review on the rolling 
tachyon:
A.~Sen, ``Tachyon dynamics in open string theory",
[hep-th/0410103].








 
\bibitem{rollingmore} 
 N. Lambert, H. Liu and J. Maldacena, ``Closed strings from
decaying D-branes" [hep-th/0303139];
P.~Mukhopadhyay and A.~Sen, ``Decay of
unstable D-branes with electric field,''  JHEP {\bf 0211}, 047
(2002)  [hep-th/0208142];
F.~Larsen, A.~Naqvi and S.~Terashima,
``Rolling tachyons and decaying branes", JHEP {\bf 0302} (2003) 039
[hep-th/0212248];
G. W. Gibbons, K. Hori and P. Yi,
``String fluid  from unstable D-branes", Nucl. Phys. {\bf B596} (2001) 136
 [hep-th/0009061];
 T. Okuda and S. Sugimoto, ``Coupling of rolling 
tachyon to closed strings", Nucl. Phys. {\bf B647} (2002) 101 
[hep-th/0208196];
C. Kim, H. B. Kim and Y. Kim, ``Rolling tachyons in 
string cosmology", Phys. Lett. {\bf B552} (2003) 111 [hep-th/0210101];
H. Lee and W. S. l'Yi, ``Time evolution of rolling 
tachyons for a brane-anti-brane pair", J. Korean Phys. Soc. {\bf 43} 
(2003) 676 [hep-th/0210221];
S.~J.~Rey and
S.~Sugimoto,  ``Rolling tachyon with electric and magnetic fields:
T-duality approach",  [hep-th/0301049];
S.~J.~Rey and S.~Sugimoto, ``Rolling of modulated tachyon
with gauge flux and emergent fundamental string", 
Phys. Rev. {\bf D68} (2003) 026003 [hep-th/0303133];
T. Takayanagi and N. Toumbas, ``A matrix model
dual of type 0B  string theory in two dimensions" [hep-th/0307083];
M. R. Douglas, I. R. Klebanov, D. Kutasov, J.
Maldacena, E.  Martinec, and N. Seiberg, ``A new hat for the $c=1$
matric model"  [hep-th/0307195];
I. Ya. Aref'eva, L. V. Joukovshaya and A. S. 
Koshelev, ``Time evolution in superstring field theory on non-BPS 
brane, JHEP {\bf 0309} (2003) 012 [hep-th/0301137];
Y. Demasure and R. A. Janik, ``Backreaction and 
the rolling tachyon", Phys. Lett. {\bf B578} (2004) 195 [hep-th/0305191];
M.~Gutperle and A.~Strominger,
``Timelike boundary Liouville theory,''  [hep-th/0301038];
F. Larsen, A. Naqvi and S. Terashima, ``Rolling 
tachyons and decaying branes", JHEP {\bf 0302} (2003) 039 
[hep-th/0212248];
N. R. Constable and F. Larsen, ``The rolling 
tachyon as a matrix model", JHEP {\bf 0306} (2003) 017 [hep-th/0305177];
V. Schomerus, ``Rolling tachyons from 
Liouville theory", JHEP {\bf 0311} (2003) 043 [hep-th/0306026];
K. Nagami, ``Rolling tachyon with electromagnetic field in linear 
dilaton background", Phys. Lett. {\bf B591} (2004) 187 [hep-th/0312149];
A. Fotopoulos and A. A. Tseytlin, ``On open 
superstring partition function in inhomogeneous rolling tachyon
background", JHEP {\bf 0312} (2003) 025 [hep-th/0310253];
E. Coletti, I. Sigalov, W. Taylor,
``Taming the Tachyon in Cubic String Field Theory",
JHEP {\bf 0508} (2005) 104 [hep-th/0505031];
V. Forini, G. Grignani, G. Nardelli,
``A new rolling tachyon solution of cubic string field theory"
[hep-th/0502151]; ``A solution to the 4-tachyon off-shell amplitude in 
cubic string field theory", JHEP {\bf 0604} (2006) 053
[hep-th/0603206]; T. Lee, ``The final facte of the rolling tachyon", JHEP {\bf 11} (2006) 056.



\bibitem{caldeira83phy}
A. O. Caldeira and A. J. Leggett,
Physica {\bf 121A}, 587 (1983).

\bibitem{fisher}
M.P.A.~Fisher and W.~Zwerger, 
Phys. Rev. {\bf B32}, 6190 (1985).


\bibitem{Tlee:2005ge}
T.~Lee and G.~W.~Semenoff,
JHEP {\bf 0505}, 072 (2005); 

\bibitem{Hassel} 
M. Hasselfield, T. Lee, G. W. Semenoff, P. C. E. Stamp,
Ann. Phys. {\bf 321}, 2849 (2006).


\bibitem{furusaki}
A. Furusaki and N. Nagaosa,
Phys. Rev. {\bf B47}, 4631 (1993).


\bibitem{tlee08}
T. Lee,
JHEP {\bf 02} 090 (2008).


\bibitem{kane92}
C. L. Kane and M. P. A. Fisher, Phys. Rev. Lett. {\bf 68}, 1220 (1992);
Phys. Rev. {\bf B46}, 15233 (1992)

\bibitem{yi}
H. Yi and C. L. Kane, Phys. Rev. {\bf B57}, R5579 (1998); H. Yi, 
``Resonant tunneling and the multichannel Kondo problem: the quantum Brownian
motion description", 
[arXiv:cond-mat/9912452].

\bibitem{affleck00}
I. Affleck, M. Oshikawa and H. Saleur, 
Nucl. Phys. B {\bf 594}, 535 (2001).


\bibitem{senechal}
D. Senechal, 
``An introduction to bosonization,"
[arXiv:cond-mat/9908262].

\bibitem{caldeira83ann} 
A. O. Caldeira and A. J. Leggett, 
Ann. Phys. {\bf 149}, 374 (1983). 





\bibitem{Affleck:1990by}
  I.~Affleck and A.~W.~W.~Ludwig,
  Nucl.\ Phys.\ B {\bf 352}, 849 (1991).

\bibitem{Oshikawa:2005fh}
 C.~Chamon,  M.~Oshikawa and I.~Affleck,
  Phys.\ Rev.\ Lett.\  {\bf 91}, 206403 (2003).


\bibitem{Thirring}
 W.~Thirring,
 Ann.\ Phys.\ {\bf 3}, 91 (1958).
 
\bibitem{Klaiber}
 B. Klaiber, in {\it Lectures in Theoretical Physics}, edited by A.~Barut and
 W.~Brittin, p. 141-176 (Gordon and Breach, New York, 1968).



\bibitem{fpl1}
N. Yu. Reshetikhin, J. Phys. A {\bf 24}, 2387 (1991).

\bibitem{fpl2}
H.W.J. Blo\"'te and B. Nienhuis, Phys. Rev. Lett. {\bf 72}, 1372 (1994).

\bibitem{fpl3}
J. Kondev, J. Gier and B. Nienhuis, J. Phys. A {\bf 29}, 6489 (1996).

\bibitem{referee}
I would like to thank the referee for bringing my attention to refs.\cite{fpl1,fpl2,fpl3}.
 
\end{thebibliography}
\end{document}